\documentstyle{xray-wuerzburg-95}

\input psfig.tex

\def\gsim{\lower 2pt \hbox{$\, \buildrel {\scriptstyle >}\over
         {\scriptstyle \sim}\,$}}
\def\lsim{\lower 2pt \hbox{$\, \buildrel {\scriptstyle <}\over
         {\scriptstyle \sim}\,$}}
\newcommand{\Msol}{\mbox{$M_{\odot}\;$}}

\begin{document}
 
\title{Temperature, distance and cooling of the Vela pulsar}
\author{Dany Page\inst{1}, Yu.~A. Shibanov\inst{2} \and V.~E. Zavlin\inst{3}}
\institute{Instituto de Astronom\'{\i}a, UNAM, M\'{e}xico D.F., M\'{e}xico;
           page@astroscu.unam.mx
           \and
           Ioffe Institute of Physics and Technology, St Petersburg, Russia;
           shib@ammp.pti.spb.su
           \and
           Max Planck Institut f\"{u}r Extraterrestrische Physik, Garching,
           Germany;
           zavlin@rosat.mpe-garching.mpg.de
           \vspace{-10.mm}}
\maketitle

\begin{figure}
\vspace{-5.5cm}
\psfig{figure=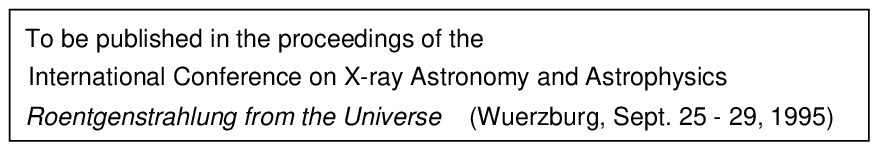}
\vspace{4.4cm}
\end{figure}
\vspace{-1.4cm}

\begin{abstract}
We use models of magnetized hydrogen atmospheres to fit the 
soft X-ray thermal spectrum of the Vela pulsar obtained by {\em ROSAT}.
The distance and hydrogen column density required by our fits are in
good agreement with other independents estimates and can be considered
as an argument for the presence of hydrogen at the surface of this pulsar.

The low temperature obtained, $T_e^\infty = 7.85 \pm 0.25 \times 10^5$~K,
is below the predictions of the `standard' model of neutron star
cooling and strengthens previous claims for the necessity of the
occurrence of fast neutrino cooling in this neutron star.

\vspace{-2.mm}

\end{abstract}

\vspace{-3.mm}
\section{INTRODUCTION}
\vspace{-2.mm}

The Vela pulsar is the youngest pulsar for which there is strong evidence
that thermal emission from the surface of the neutron star has been detected
(\"Ogelman, Finley \& Zimmermann 1993; \"Ogelman 1995).
A reliable measurement of its surface temperature based on modeling of
the surface thermal emission is of outmost importance for comparison
with models of neutron star cooling and could give us evidence for the past 
occurrence of fast neutrino emission.
Moreover, since the chemical composition of a pulsar's surface and its state
(gaseous, liquid or solid) are still unknown, the `successful' fit of the
observations can also give us invaluable information about
this question.

\vspace{-2.mm}
\section{MAGNETIZED HYDROGEN ATMOSPHERE}
\vspace{-2.mm}

There are many possibilities for the structure of a pulsar's surface and a
magnetized hydrogen atmosphere is the most `natural' and simplest choice
although it has {\em a priori} no strict reason to be accepted 
(e.g., where does the hydrogen come from ?).
Assuming the presence of an atmosphere, the magnetic field has three
main effects: 
1) the opacity depends strongly on the magnetic field direction and on 
the photon polarization,
2) for the polarization giving the main contribution to the emergent 
flux it is much smaller than in the nonmagnetic case 
and its dependence on the photon energy is weaker,
3) the ionization energy is strongly increased and at low $T_e$ an absorption
edge appears within the range of the {\em ROSAT}$\,$ PSPC.
The problem of ionization equilibrium in a magnetized hydrogen atmosphere
is not yet completely solved but at $T_e$~$\gsim$~$10^6$~K the whole atmosphere
is fully ionized, so its structure can be calculated accurately and
the spectra we are using are reliable
(Shibanov $et~al$. 1992; Pavlov $et~al$. 1994, 1995).

\vspace{-2.mm}
\section{SPECTRAL FITS}
\vspace{-2.mm}

We performed a $\chi^2$ search of the whole parameter space
using the {\em ROSAT} data as presented by 
\"Ogelman, Finley \& Zimmermann (1993),
restricting ourselves to $E < 1.5$ keV to eliminate the hard tail.
Due to the high temperature, several simplifications occur when considering
emission within the {\em ROSAT}$\,$ PSPC band:
1) the emitted flux is weakly dependent 
on the field strength at $B\gsim 10^{12}$ G,
2) the detectable flux practically depends only on the mean `effective' 
surface temperature, not on the actual distribution of the temperature 
along the surface (Page 1995a, Page \& Sarmiento 1995),
3) the red-shift is equivalent to a change (red-shift) of temperature and
4) the model depends only weakly on the surface gravity.

As a first approximation we are thus able to do spectral fits with only
three parameters:
effective temperature at infinity, $T_e^{\infty}$,
distance $D$ and column density $N_H$.
We assume a mass of 1.4 \Msol and a radius of $R = 10$ km, i.e., 
$R^{\infty} = 13.6$ km, but since the received flux scales like 
$(R^{\infty}/D)^2$ our results can be extrapolated to 
other radii with reasonable confidence.
The final result of our $\chi^2$ search is shown in Figure 1.
%

The pulsations of the soft X-ray emission from Vela  show that the
temperature of the polar cap region may be even larger than the mean
effective surface tempeature found in our fit 
($T_{s\!u\!r\!f} \sim 9 \times 10^5$ K ) 
and the fully ionized atmosphere
fit might be quite reliable for this object.
However, at $T_{s\!u\!r\!f} \lsim 10^6$ K an absorption edge (not taken into
account in the spectra we used) should appear: as a result our fits give too low
a $T_e$ and our lower range would probably be pushed to higher values.

\begin{figure}
\vspace{-4.mm}
\psfig{figure=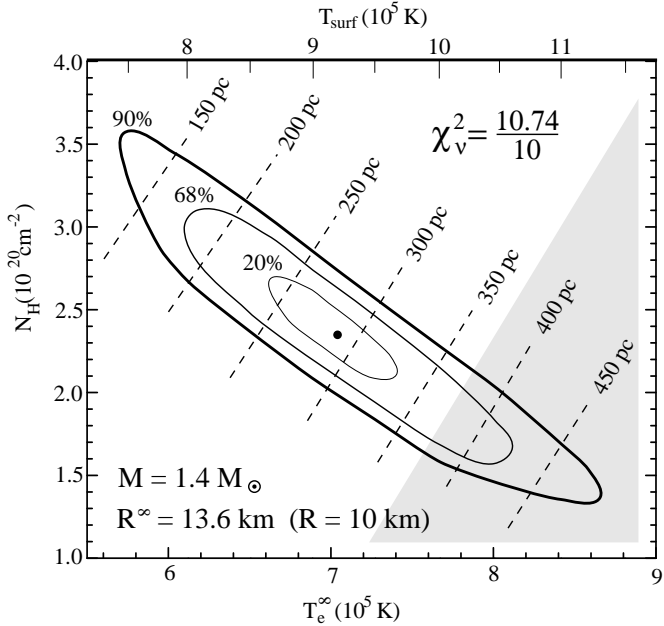}
\vspace{-3.mm}
\caption[]{Confidence contours for the three parameters ($T_e$, $N_H$ and $D$),
           projected onto the $T_e - N_H$ plane.
           The grey area shows the `$1 \sigma$' lower limit on the distance
           from the pulsar dispersion measure.
           The dot shows our best fit: 
           $T_e^{\infty} = 7.04 \times 10^5$ K, 
           ($T_{s\!u\!r\!f} = 9.2\times 10^5$ K), 
           $N_H = 2.35 \times 10^{20} {\rm cm}^{-2}$ and
           $D = 285$ pc.}
\vspace{-2.0mm}
\end{figure}

\vspace{-2.mm}
\section{DISCUSSION}
\vspace{-2.mm}

We obtain in our best fit a $\chi^2$ per degree of freedom of 1.074;
due to the low statistics any $\chi^2$ p.d.f. of order one is acceptable.
So the relevance of the model can only be assessed by discussion of
the resulting values of the parameters.

\noindent
{\it{Column density:
N$_H$ = 2.35 $\pm$ 0.8 $\times$ 10$^{20}$ cm$^{-2}$}.} 
\noindent
Other components of the Vela PSR + SNR give independent estimates of $N_H$.
Modeling of the pulsar `hard tail' soft X-ray emission needs values of
4.3~$\pm$~1.0 or 4.4~$\pm$~1.5~$\times 10^{20}$ cm$^{-2}$ and 
modeling of the surrounding compact nebula gives values of 2.0~$\pm$~0.5
or 3.0~$\pm$~0.5~$\times 10^{20}$ cm$^{-2}$ (\"Ogelman {\em et al.} 1993)
while the six `bullets' detected in the SNR give $N_H$'s between 
2.4~$\pm$~1.9 and 5.4~$\pm$~2.1~$\times~10^{20}$~cm$^{-2}$ 
(Aschenbach {\em et al.} 1995).  
In addition, Table 1 list several reference objects in the line of sight of
the Vela pulsar for comparison with our results which show that $N_H$
reaches $10^{21}$~cm$^{-2}$ well above 500~pc and probably needs
at least 100~--~200 pc to reach $10^{20}$~cm$^{-2}$.
In short, a $N_H$ between 1.5 and 3~$\times~10^{20}$~cm$^{-2}$ seems
reasonable.

\noindent
{\it{Distance:
            D = 300 $\pm$ 120 pc}.} 
The pulsar's distance of 500 $\pm$ 125 pc (Taylor {\em et al.} 1993),
confirmed by interstellar scintillation measurements (Gupta 1995), is
compatible, in its lower range, with our deduced value, in its upper range.
Moreover, increasing the star's radius $R^{\infty}$ increases the fitted
distance $D$ in the same amount. 
Our results may possibly favor a larger radius: $R= 12$ km 
($R^{\infty} = 15.5$ km) would increase $D$ by 14\%.

Our 68\% confidence range (`$1 \sigma$') combined with the pulsar's distance 
(at `$1 \sigma$', the grey area in Fig. 1) thus implies that 
$T_e^\infty  = 7.85 \pm 0.25 \times 10^5$K (for a 10 km radius)
and $N_H = 1.80 \pm 0.25 \times 10^{20}$ cm$^{-2}$.
%
The corresponding surface temperature $T_{s\!u\!r\!f}$ is above
10$^6$ K where our atmosphere models are most reliable.

\noindent
{\it{Hydrogen atmosphere ? }} 
The consistency of our result for $D$ and $N_H$ with values obtained with
other methods is an argument in favor of the presence of a hot ($\sim10^6$ K) 
magnetized ($\gsim 10^{12}$ G) hydrogen plasma at the surface of 
the Vela pulsar in sufficient amount to provide an optical
depth of unity, i.e., at least a few grams per cm$^2$. 

\begin{table}[t]
\vspace{-3.mm}
\caption{Reference objects for interstellar absorption \newline
\label{tab:CEA}}
\begin{tabular*}{8.8cm}{ccccc}
\hline
Name (HD number)     & $l$    & $b$  &  $\begin{array}{c}
                                   \rm Log N_H \\ \rm [cm^{-2}]
                                        \end{array}$       & $\begin{array}{c}
                                                             \rm D \\ \rm [pc]
                                                             \end{array}$  \\
\hline
IX Vel$^a$           & 264.9 & -7.9 &     19.30            &       140     \\
(HD72350)$^a$        & 262.7 & -3.2 &        $<$ 21.15     &       292        \\
\bf Vela PSR$^b$     & 263.6 & -2.8 &     20.35 $\pm$ 0.15 
                                                  & 300 {\scriptsize$\pm$ 120}\\
  Vela PSR$^c$       & 263.6 & -2.8 &     19 -- 20.2       & $\sim 1,500$     \\
HX Vel (HD74455)$^a$ & 266.6 & -3.6 &     20.78 $\pm$ 0.10 &       422        \\
(HD72179)$^a$        & 262.1 & -3.0 &        $<$ 21.32     &       607        \\
KX Vel (HD75821)$^a$ & 266.3 & -1.5 &     20.48 $\pm$ 0.10 &       972        \\
(HD74531)$^a$        & 266.7 & -3.6 &     20.84 $\pm$ 0.10 &       996        \\
(HD73658)$^a$        & 264.7 & -3.1 &     21.20 $\pm$ 0.05 &      1508        \\
\hline
\end{tabular*}

\baselineskip=9pt
\begin{scriptsize}
$a$: Fruscione, A. {\em et al.} 1994. \newline
$b$: This work \newline
$c$:  \"Ogelman, Finley, Zimmermann 1993. $T_e^{\infty} \sim 1.6 \times 10^6$ K, 
blackbody fit.
\end{scriptsize}

\vspace{-3.mm}

\end{table}

\vspace{-2.mm}
\section{COOLING}
\vspace{-2.mm}

The Vela pulsar has already been claimed several times to be cooler
than the predictions of the `standard' model of neutron star cooling
and our results reinforce such claims.  If we accept our deduced $T_e$
and believe neutron star cooling models then our results require fast
neutrino emission by, e.g., direct Urca processes, kaon or pion
condensates, which has been supressed by baryon pairing (e.g., Page \&
Applegate 1992).  A critical temperature $T_c$ of the order of $1 - 3
\times 10^9$ K for pairing in the inner core is then necessary,
independently of the exact fast cooling agent (Page 1995b).

\vspace{2.mm}

\baselineskip=6pt
\begin{tiny}
\noindent
The work of D.P. was supported by grants IN 105794 and IN 105495
from UNAM-DGAPA.
The work of Yu.A.S. and V.E.Z. was partly supported
by INTAS grant 94-3834 and RFFI grant 93-02-2916.
\end{tiny}

\vspace{-2.mm}

 

\begin{thebibliography}{}

\bibitem{}
Aschenbach B., Egger R., Tr\mbox{\"{u}}mper J., 1995, 
Nature 373, 587

\bibitem{}
Fruscione A., Hawkins I., Jelinsky P., Wiercigroch, A., 1994, 
ApJS 94, 127

\bibitem{}
Gupta Y., 1995,
ApJ 451, 717

\bibitem{}
\mbox{\"{O}}gelman H., 1995. In {\em The Lives of the Neutron Stars},
eds. A. Alpar, \mbox{\"{U}}. Kizilo\mbox{\u{g}}lu, \& J. van Paradijs
(Dordrecht: Kluwer Academic Publishers), 101

\bibitem{}
\mbox{\"{O}}gelman H., Finley J.~P., Zimmermann H.-U., 1993,
Nature 361, 136

\bibitem{}
Page D., 1995a, 
ApJ 442, 273
(astro-ph/9407015)

\bibitem{}
Page D., 1995b, 
Rev. Mex. F\mbox{\'{\i}}s. 41, Supl. 1, 178
(astro-ph/9501071)

\bibitem{}
Page D., Applegate J. H., 1992, 
ApJ 394, L17

\bibitem{}
Page D., Sarmiento, A., 1995, these proceedings.
(astro-ph/9601188)

\bibitem{}
Pavlov G.~G., Shibanov Yu.~A., Ventura J., Zavlin V.~E., 1994,
A\&A 289, 837

\bibitem{}
Pavlov G.~G, Shibanov Yu.~A., Zavlin V.~E., Meyer R., 1995.
In {\em The Lives of the Neutron Stars},
eds. A. Alpar, \mbox{\"{U}}. Kizilo\mbox{\u{g}}lu, \& J. van Paradijs
(Dordrecht: Kluwer Academic Publishers), 71

\bibitem{}
Shibanov Yu.~A., Zavlin V.~E., Pavlov G.~G, Ventura J., 1992,
A\&A 266, 313

\bibitem{}
Taylor J. H., Manchester R. N., Lyne A. G., 1993,
ApJS 88, 529


\end{thebibliography}
\end {document}